\documentclass[runningheads]{llncs}
\usepackage{graphicx}
\usepackage{multirow}
\usepackage{latexsym}
\usepackage{caption}
\usepackage{xcolor}
\usepackage{amsmath,amsfonts,amssymb}
\usepackage[thinlines]{easytable}
\usepackage[colorlinks=true, allcolors=blue]{hyperref}
\usepackage{subcaption}
\usepackage[style=numeric,sorting=none]{biblatex}
\begin{document}
\title{Applying Machine Learning Models on Metrology Data for Predicting Device Electrical Performance}
\titlerunning{Machine Learning for Predicting Device Electrical Performance from Metrology Data}
%
\author{Bappaditya Dey\inst{1}* \and Anh Tuan Ngo\inst{2}* \and Sara Sacchi\inst{1 \and 3}* \and Victor Blanco\inst{1} \and Philippe Leray\inst{1} \and Sandip Halder\inst{1}}

\institute{imec, Kapeldreef 75, 3001 Leuven, Belgium \and School of Engineering and Physical Sciences, Henriot-Watt University, Edinburgh (UK) \and Department of Physics and Astronomy, University of Bologna, Italy}

\maketitle              
\vspace*{-0.59cm}
\begin{center}
    \small * Bappaditya Dey, Anh Tuan Ngo and Sara Sacchi contributed equally to the work.
\end{center}

\begin{abstract}
Moore’s Law states that transistor density will double every two years, which is sustained until today due to continuous multidirectional innovations (such as extreme ultraviolet lithography, novel patterning techniques etc.), leading the semiconductor industry towards 3 nm node (N3) and beyond. For any patterning scheme, the most important metric to evaluate the quality of printed patterns is edge placement error, with overlay being its largest contribution. Overlay errors can lead to fatal failures of IC devices such as short circuits or broken connections in terms of pattern-to-pattern electrical contacts. Therefore, it is essential to develop effective overlay analysis and control techniques to ensure good functionality of fabricated semiconductor devices. In this work we have used an imec N-14 BEOL process flow using litho-etch-litho-etch (LELE) patterning technique to print metal layers with minimum pitch of 48nm with 193i lithography. Fork-fork structures are decomposed into two mask layers (M1A and M1B) and then the LELE flow is carried out to make the final patterns. Since a single M1 layer is decomposed into two masks, control of overlay between the two masks is critical. The goal of this work is of two-fold as, (1) to quantify the impact of overlay on capacitance and (2) to see if we can predict the final capacitance measurements with selected machine learning models at an early stage. To do so, scatterometry spectra are collected on these electrical test structures at (a) post litho, (b) post TiN hardmask etch, and (c) post Cu plating and CMP. Critical Dimension (CD) and overlay measurements for line/space (L/S) pattern are done with SEM post litho, post etch and post Cu CMP. Various machine learning models are applied to do the capacitance prediction with multiple metrology inputs at different steps of wafer processing. Finally, we demonstrate that by using appropriate machine learning models we are able to do better prediction of electrical results.

\end{abstract}
\section{Introduction}
\begin{figure}[h!]
\includegraphics[width=6cm]{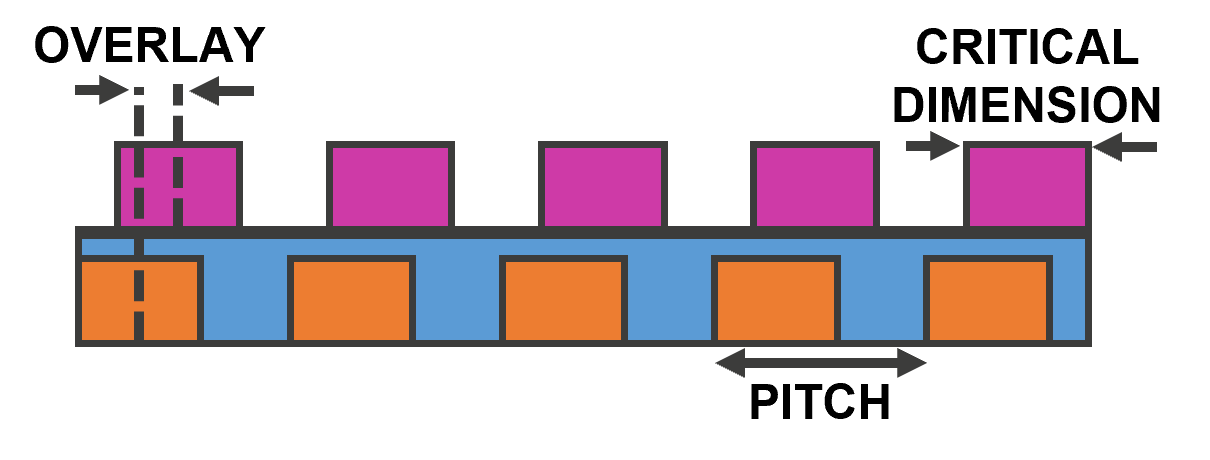}
\centering
\vspace{3mm}
\caption{Illustration of overlay, critical dimension, and pitch.} \label{fig111}
\end{figure}

The long standing IC industry push for device shrink, increased drive current and lower operating voltages often results in complex 3D device architectures. The inspection of 3D architectures imposes more challenges and demands in increased importance of metrology \cite{ref_1}. During R\&D phase, different metrology techniques are used for exploratory process development, while, during high volume manufacturing, metrology is focused on process control. Techniques like scatterometry and CD-SEM (Critical Dimension Scanning Electron Microscope) are typically used for in-line CD measurements. However, both techniques have certain limitations and advantages. For instance, inconsistency in material properties (n\&k) and long model-optimization times restrict scatterometry techniques, while resist shrinkage and its charging effect impact the measurement performance of the CD-SEM tool \cite{ref_2}. As advanced patterning techniques involve fabricating the device in multiple steps of different layers, critical dimensions and overlay are crucial to be monitored and controlled. Overlay can be defined as the relative alignment of consecutive masked layers within the device \cite{ref12}. Fig. \ref{fig111} shows an illustration of overlay, critical dimension, and pitch of devices on a wafer. If the overlay errors between layers exceed an allowed budget, defects such as short circuit, line break, or bad contacts will occur, which leads to fatal failure of the devices. 

To ensure the quality of fabricated devices in multiple patterning schemes, a number of metrologies are conducted after each process step. This leads to higher costs and lower yield in the production line. One solution to tackle this problem is to build mathematical models that can predict the properties of the devices in the last steps (critical dimensions or electrical performance) using the metrology data from early steps. With these mathematical models, a number of intermediate metrology steps can be reduced, thus increasing the throughput. Furthermore, for high-volume manufacturing, gaining an early insight into the qualities of final structures can be a game changer, since proper actions can be made in timely manner to improve the overall process. For example, if a model using the overlay data measured after the second lithography step can predict that there will be a lot of failures on the wafer in the final steps, the wafer can be reworked, and thus can significantly save resources and cost of production. This will not be possible if we only discover faulty wafers in later steps, where these wafers can only be scrapped.

Machine learning solutions have become an attractive tool for future process control and monitoring purposes \cite{ref5}. A systematic survey was conducted on recent research works, which demonstrates different machine learning/deep learning techniques applied towards improving EPE in semiconductor manufacturing domain \cite{ref1000}. The methodology deployed in this study involves using the overlay metrology data from early process steps. Then, a mathematical estimator is generated using a set of machine learning algorithms. The data used to create the mathematical estimator is the training set. Once the training was completed, showing a good correlation to the reference data, the overlay metrology data from early process steps was used to predict different outcomes, such as CD and electrical performance (say, capacitance of final fabricated structures) for other wafers \cite{ref_6}. This technique is less dependent on structural complexity. The early estimation of electrical performance and variability provided by these machine learning techniques can significantly reduce cost for both R\&D and high-volume manufacturing (HVM), by taking proper action in a timely manner either to scrap or rework the wafer, and improve or monitor the process. 

\section{Experimental Design and Methods}
\subsection{Description of the process flow - LELE approach} 

\begin{figure}[h!]
\includegraphics[width=8cm]{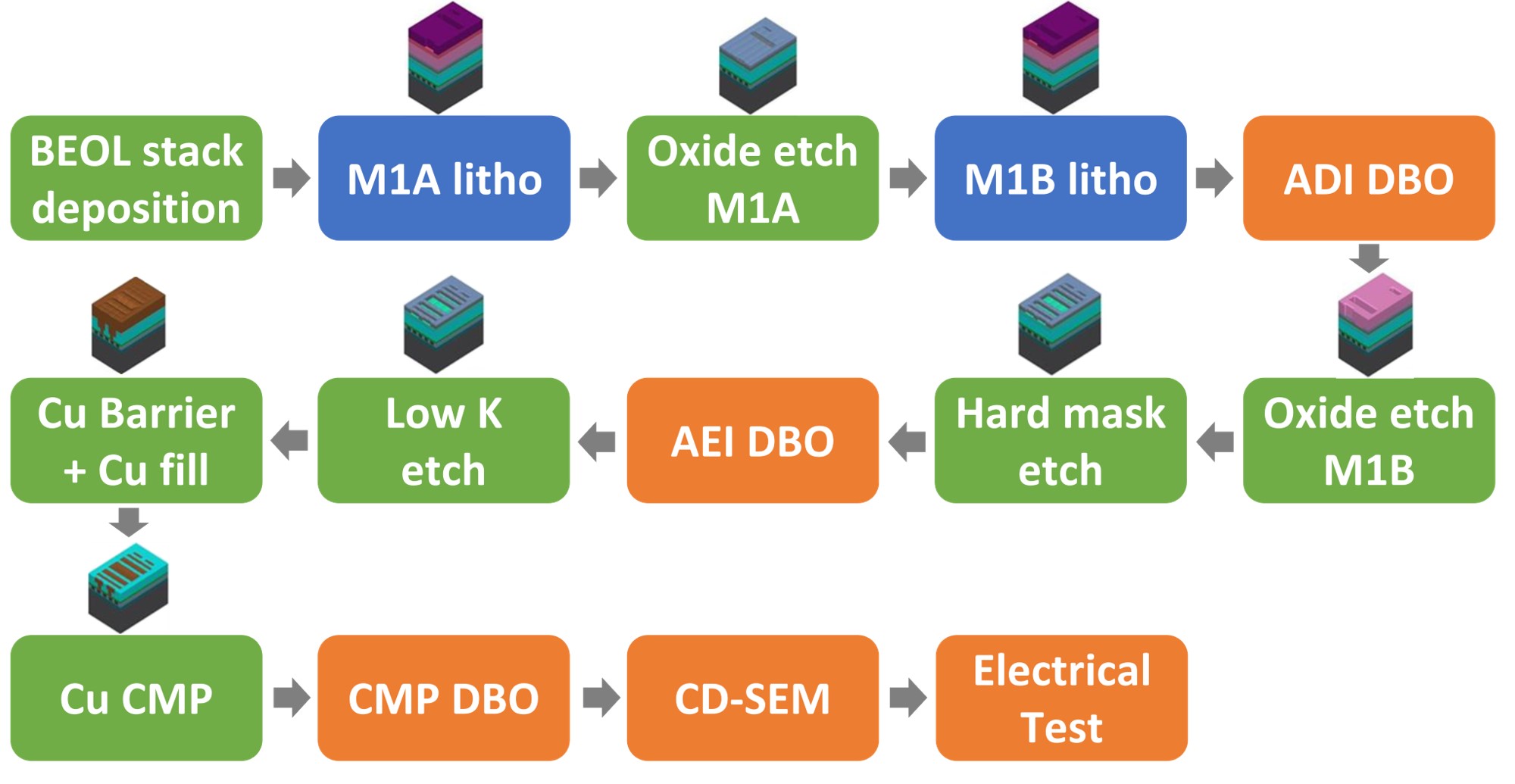}
\centering
\vspace{3mm}
\caption{Process flow showing LELE approach to achieve 48 nm pitch.} \label{fig1}
\end{figure}

The studied layer uses a Litho-Etch-Litho-Etch (LELE) process flow to pattern a 48 nm M1 pitch by means of a double exposure of M1A and M1B layers, each of them at 96 nm pitch using a 193 nm immersion scanner. A simplified process flow describing the main fabrication steps is shown in Fig. \ref{fig1}. It starts with Back-End-Of-Line (BEOL) stack deposition followed by M1A lithography exposure using an ASML NXT 1950i immersion scanner. The post litho trench target CD was 40 nm. A negative tone lithography development (NTD) process was chosen to pattern the trenches. An 85  nm resist was coated on top of 30 nm (spin on glass) SOG and 100 nm (spin on carbon) SOC. This was used to pattern the M1A layer which was then transferred onto an oxide pattern storage layer by etching in a plasma chamber. The oxide etch step involved opening the SOG/SOC layer and pattern transfer into the oxide layer. A second lithography exposure was performed to pattern the M1B layer using same lithography and etch process as for M1A layer.. At this stage ADI (After Develop Inspection) overlay between M1A and M1B was measured using Diffraction-Based Overlay (DBO) targets. This was followed by transferring the pattern onto the same oxide pattern storage layer. After that the patterns from M1A and M1B were transferred onto a TiN hardmask (HM) layer by plasma etching. At this stage, CD data using CD-SEM, AEI (After Etch Inspection) overlay data, as well as scatterometry spectra were collected on electrically active device structures. Finally, the TiN HM layer was used to transfer the pattern into the low k dielectric material to create the 48 nm pitch M1 trenches. Trenches were filled with Cu, followed by a CMP (chemical mechanical polishing) step. A 5 nm thin layer of SiCN layer was also deposited on top of the wafer to prevent oxidation of Cu metal lines. Scatterometry spectra collection were done on all resistance and capacitance targets before they go for electrical tests.

\subsection{Description of the electrical test structures}
\begin{figure}[h!]
\hspace*{-0.50in}
\includegraphics[width=10cm]{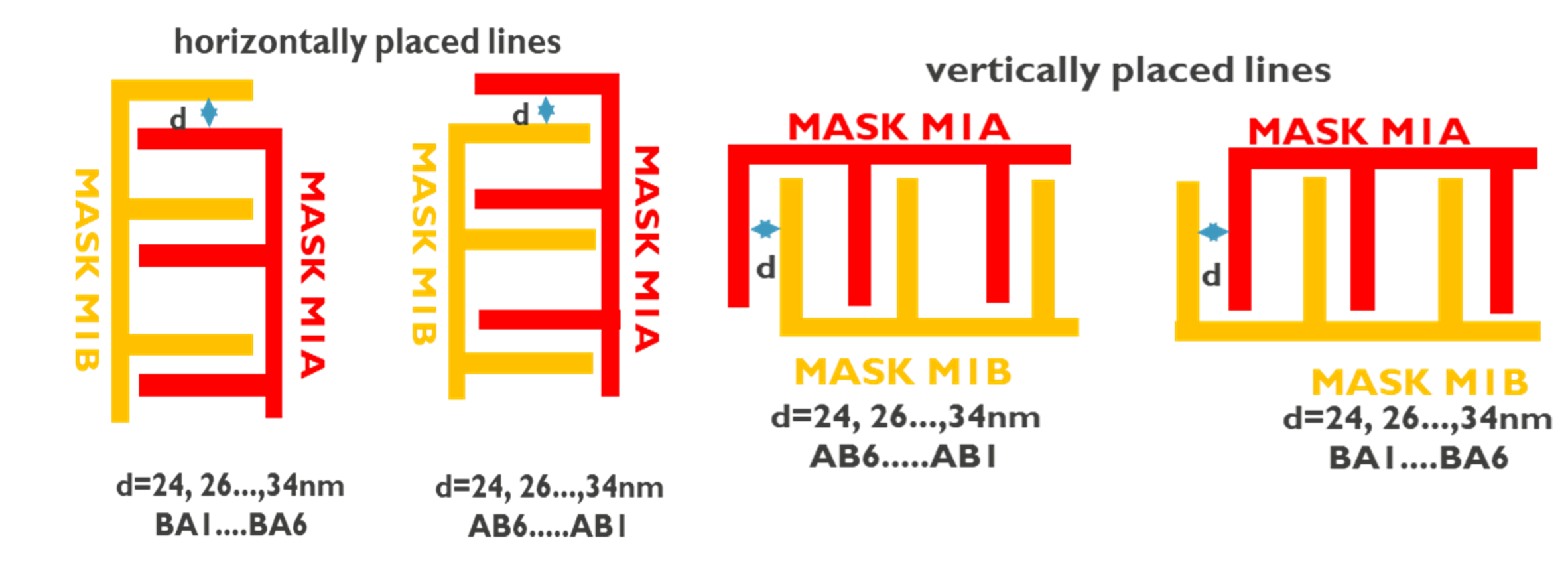}
\centering
\vspace{3mm}
\caption{Description of capacitance measurement fork-fork structures.}
\label{fig3}
\end{figure}

Fig. \ref{fig3} describes the 12 vertically placed and 12 horizontally placed (in the layout) fork-fork structures to measure capacitance. Due to overlay errors, the dielectric distance (d) between M1A and M1B changes and can be directly correlated with X overlay for vertical lines and with Y overlay values for horizontal lines. The design critical dimension here is 24 nm and the distance between the two metal lines changes in steps of 2 nm. For example, the distance between M1A and M1B is 24 nm for the AB6 and 34  nm for AB1. Similarly, it is 24 nm for BA1 and 34 nm for BA6. 

\subsection{CD and overlay fingerprints post TiN HM etch}
\begin{figure}[h!]
\includegraphics[width=9cm]{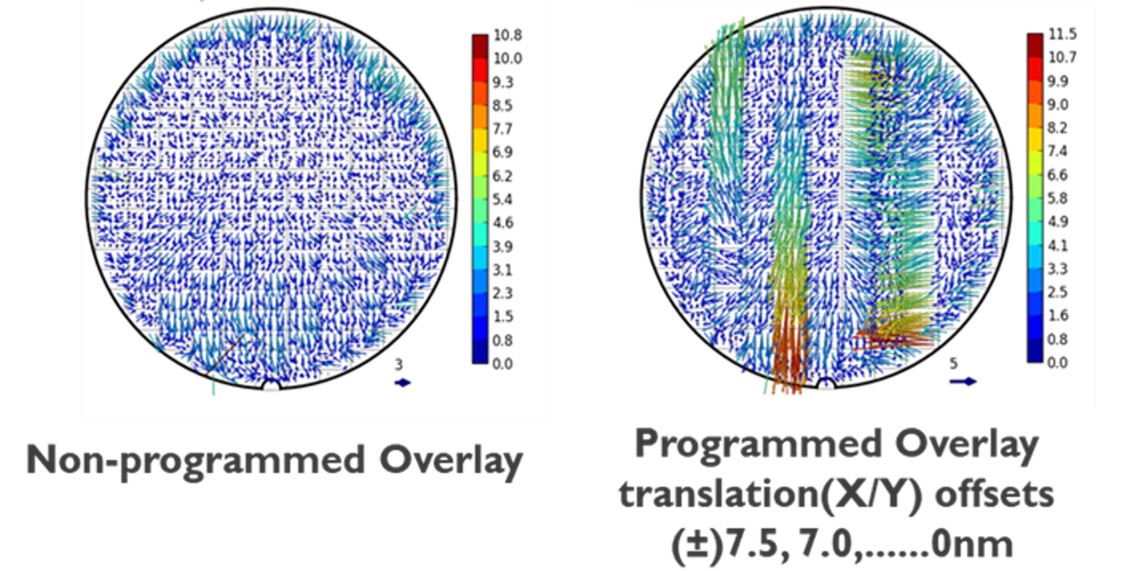}
\centering
\vspace{3mm}
\caption{Example of ADI overlay fingerprint for a POR wafer and wafer exposed with programmed overlay.}
\label{fig4}
\end{figure}

In order to produce a robust design of experiment we varied different parameters both at litho and etch steps. Wafers with programmed overlay were fabricated by creating a scanner sub-recipe such that wafers receive a translational offset of 0 to ±7.5 nm in X and Y direction on selected dies as shown in Fig. \ref{fig4}. 

\subsection{Diffraction Based Overlay - DBO}
\begin{figure}
\centering
    \subfloat[Normal incidence reflection]{\includegraphics[width=3cm]{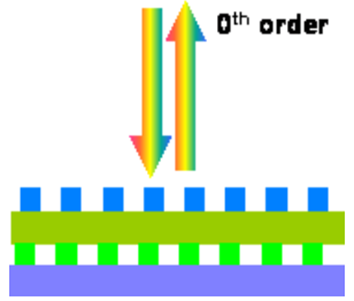} \label{fig15} }
    \qquad
    \subfloat[DBO target pads]{\includegraphics[width=7cm]{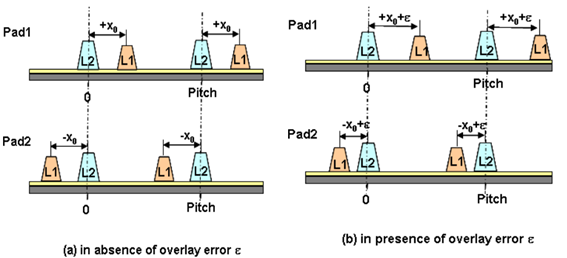} \label{fig16} }
    \caption[Schematic diagrams of normal incidence reflection and DBO target pads.]{Schematic diagrams illustrating (a) a normal incidence reflection and (b) the DBO target designed pads \cite{ref9}.}
    \label{fig150}
\end{figure}

The LELE double patterning technique not only entails more complex process stages, but it also needs tighter overlay control compared to the conventional single pattering \cite{ref9}. Thus, it is crucial to use advanced overlay metrology methods with very high certainty to ensure the quality of fabricated devices. Recently, Diffraction-Based Overlay (DBO) emerges as a promising overlay metrology technique which offers better precision than the conventional Image-Based Overlay method \cite{ref10}. DBO technique was used to monitor the overlay error of the LELE process. In DBO method, structures of stacked periodic gratings are used as the targets to measure overlay error. This method uses polarized light, with a broadband of wavelengths, perpendicularly projected to the grating targets and it measures the zero-order diffracted signal as a function of wavelength (Fig. \ref{fig15}). The overlay error is measured between two layers, and spectra are obtained from target pads, each of which has gratings printed in both layers. The gratings of each target pads are similar, but intentionally shifted to each other by design as shown in Fig. \ref{fig16}.

Due to symmetry, reflected spectra from pads with shifts of similar amplitude but in opposite direction are equal:

\begin{equation}
    R(+x_0,\lambda) = R(-x_0,\lambda)
\end{equation}

Here $R(x_0,\lambda)$ is the reflected spectrum from one target pad as a function of wavelength $\lambda$ and shift $x_0$. In the presence of overlay error ($\epsilon$), the symmetry is broken, and the differential spectrum of two pads is described as follows:

\begin{equation}
    \Delta R(\lambda) = R(x_0 + \epsilon,\lambda) - R(x_0 - \epsilon,\lambda) \cong 2 \epsilon \frac{\partial R}{\partial x}|_{x_0}
\end{equation}

Using another pad with known relative offset ($x_0+\delta$) to get the differential spectrum $\Delta R'(\lambda)$ with respect to the $x_0$-shifted pad (Eq. \ref{eq1}), the overlay error ($\epsilon$) can be calculated as in Eq. \ref{eq2}.

\begin{equation}
    \Delta R'(\lambda) = R(x_0 + \epsilon,\lambda) - R(x_0 - \epsilon,\lambda) \cong \delta \frac{\partial R}{\partial x}|_{x_0}
    \label{eq1}
\end{equation}

\begin{equation}
    \epsilon = \frac{\delta \Delta R(\lambda)}{2\Delta R'(\lambda)}
    \label{eq2}
\end{equation}

\section{Experimental Results and Discussions}
 A set of experiments and analysis using overlay data have been performed to find the optimized solutions for the process control in semiconductor manufacturing. The
final goal of the experiments is to predict the electrical performance of test structures from the overlay metrology data in early process steps.
Making it possible will enable/allow process engineers to make the decision of reworking or scrapping the wafers, or let them go through the next steps in the process. This will significantly help optimization of the process control, as well as save time and resources for the manufacturing operations.

\subsection{Space-CD prediction of test structures from overlay data}
Predictions of the space critical dimension (CD2) of test structures from previously measured overlay metrology data are carried out through various machine learning algorithms. The best performed model was then used to predict the capacitance value of the fabricated structures.

\subsubsection{Methodology}
In this experiment, overlay metrology and CD-SEM data of four wafers are collected. Two of them are Non-Programmed Overlay wafers, while the other two are Programmed Overlay wafers, as showed in Table \ref{tab100}. 

\vspace{-1.5mm}
\begin{table}[h!]
\centering
\caption{Description of wafers used in the experiment.}\label{tab100}
\vspace{3mm}
\begin{tabular}{|c|c|}
\hline
 {\bf Wafers} & Recipes \\
 \hline
 \bf D02 & Non-Programmed Overlay    \\
 \bf D03 & Non-Programmed Overlay \\
 \bf D10 & Programmed Overlay \\
 \bf D11 & Programmed Overlay \\
\hline
\end{tabular}
\end{table}

\begin{figure}[h!]
\includegraphics[width=10cm]{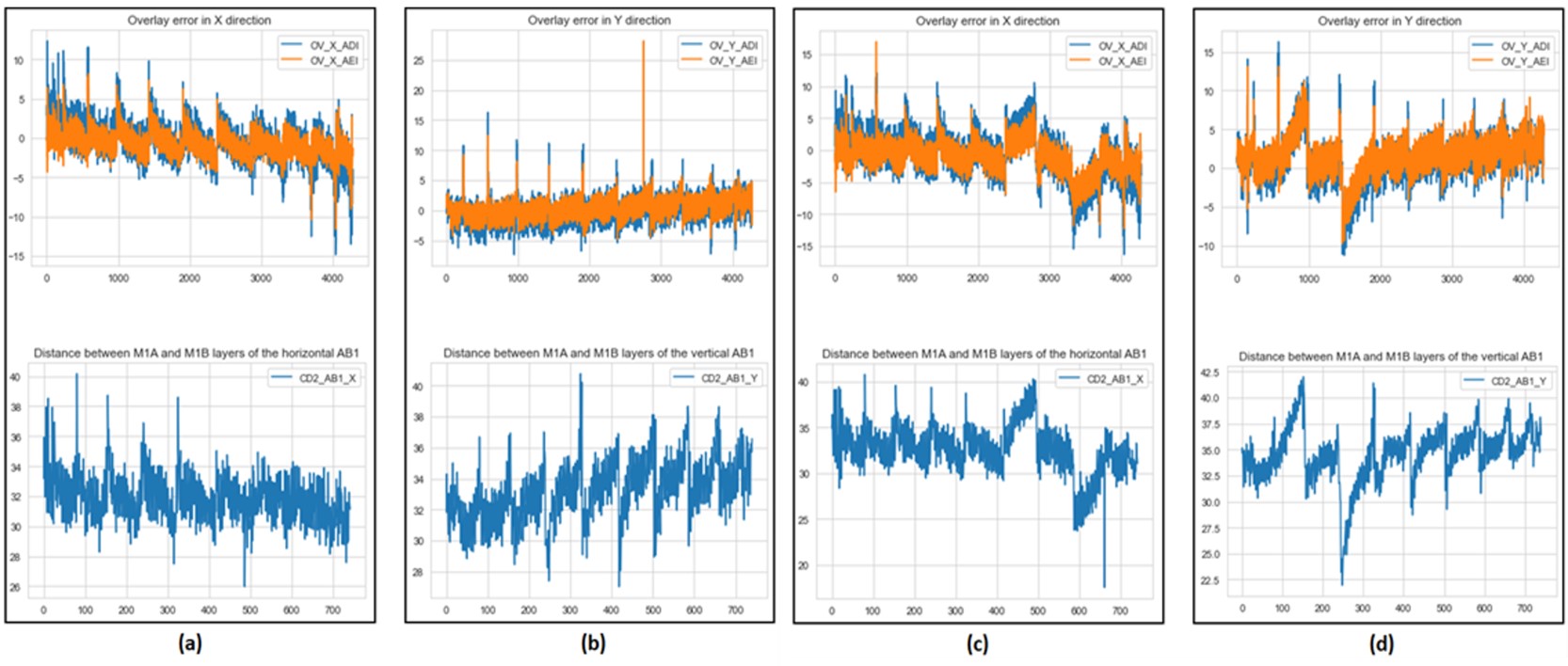}
\centering
\vspace{3mm}
\caption{Collected overlay and CD2\_AB1 data. (a) and (b) show the overlay and CD-SEM data of a Non-Programmed Overlay wafer in X and Y directions. (c) and (d) show the overlay and CD-SEM data of a Programmed Overlay wafer in X and Y directions. For all graphs, the vertical axis represents the values of overlay errors (top graph) and CD2\_AB1 (bottom graph) in nanometers, while the horizontal axis the sequence of data points.}
\label{fig52}
\end{figure}

In CD-SEM measurement, a number of metrics such as line width and distance of M1A and M1B layers or line width roughness are measured. The aim is to predict the distance (CD2) between M1A and M1B layers of the AB1 test structures (CD2\_AB1), which has the designed value of 32  nm, using overlay data from previous steps. The line width and distance of other test structures can be predicted with the same method. The ADI-AEI overlay and CD2\_AB1 data in X direction of a Non-Programmed Overlay wafer and a Programmed Overlay wafer is shown in Fig. \ref{fig52}. In this figure, the vertical axes represent the value of overlay errors (top graph) and CD2\_AB1 distances (bottom graph) in nanometers. The horizontal axes represent the sequence of data points. Qualitatively, we can see that CD2\_AB1 data aligns well with the pattern of overlay data.

Machine Learning models are built to predict the space critical dimension (CD2) of horizontal-placed and vertical-placed AB1 structure separately. The performance of those models is also compared when overlay data from different measuring steps (ADI DBO, AEI DBO, CMP DBO) are used for training and predicting. In the dataset, each of the 149 dies on the wafer was measured in 4 to 5 target points using CD-SEM, which results in more than 700 datapoints for each wafer. The CD2 of these measured points are the target labels for the Machine Learning models. Among 4 wafers that were measured in this experiment, data from three wafers (D02, D03, and D11) was used as the training set, and data from wafer D10 was used for validation.

There are in total 3 datasets, one for each DBO step (ADI, AEI, and CMP), and each one contains 30 input features:
\begin{itemize}
    \item 26 overlay measurements of each die. Overlay error is chosen in Y direction, if the target CD2\_AB1 is of the horizontal structures, and in X direction, if CD2\_AB1 is of the vertical structures.
    \item Die positions in X and Y directions.
    \item Target positions of each CD-SEM measurement on each die in X and Y.
\end{itemize}

To predict the CD2\_AB1, four Machine Learning regression algorithms were chosen to be compared, i.e. \textit{Linear Regression}, \textit{Support Vector Regression (SVR)}, \textit{Random Forest} and \textit{Extra Tree Regressor}, as they are among the most popular and powerful. These Machine Learning models were built using a Python module called Scikit-learn \cite{ref11}. For \textit{SVR} model, Radial Basis Function kernel was used, Regularization parameter was chosen to be 3, and the epsilon parameter was 0.07. For \textit{Random Forest} and \textit{Extra Tree}, the forest size and minimum sample split size were chosen to be 60 and 2, respectively. Before training, the dataset is normalized and scaled to unit variance using equation (Eq. \ref{eq5}).
\begin{equation}
    z=\frac{x}{s}
    \label{eq5}
\end{equation}
Where z is the data after being normalized, x is the training sample, and s is the standard deviation of x. Two metrics have been used to evaluate the models: the coefficient of determination ($R^2$) (Eq. \ref{eq6}), and the mean square error (MSE) (Eq. \ref{eq7}). 

\begin{equation}
    R^2=1-\frac{\sum_{i}(y_i - f_i)^2}{\sum_{i}(y_i - \overline y) }
    \label{eq6}
\end{equation}

\begin{equation}
    MSE=\frac{1}{n}\sum_{i=1}^n (y_i - f_i)^2
    \label{eq7}
\end{equation}

Here, y=[$y_{1}$,$y_{2}$,…,$y_{n}$] is the observed data, f=[$f_{1}$,$f_{2}$,…,$f_{n}$] is the corresponding fitted value of $y$ by a regression model, and $\overline y$ is the mean of $y$. The $R^2$ metric has the maximum value of 1, which means that the predicted values are exactly the same as the observed values. If the regression models always give the prediction value of $\overline y$, then $R^2$=0, which is called baseline. $R^2$ value is negative for a prediction worse than the baseline. On the other hand, MSE value is always greater than or equal to zero. The closer this metric is to zero, the better the performance of the models.

\subsubsection{Results and Discussions}
The two tables below compare the performance of four Machine Learning models using the data from the three DBO steps. Table \ref{tab111} shows the results of these model with the datasets of horizontal AB1 structures, while Table \ref{tab112} shows the results when the datasets are collected from the measurement of vertical AB1 structures. It can be seen that different algorithms resulted in different performance accuracies. In both horizontal and vertical dataset, it is expected that \textit{Linear Regression} showed the worst performance compared to the other, due to its assumption that the data only follows a linear trend, which is not the case for most datasets. \textit{SVR} model is more complex than the \textit{Linear Regression}, and thus can better capture the underlying trend of the dataset. However, the prediction accuracy of \textit{SVR} is still not as good as the \textit{Random Forest} and the \textit{Extra Tree Regressor} models for most of the cases.

\begin{table}[h!]
\centering
\caption{Evaluation results of the four ML models for the horizontal AB1 structures dataset.}\label{tab111}
\vspace{5mm}
\begin{tabular}{|c|c|c|c|c|}
\hline 
 &  & {\bfseries ADI DBO} & {\bfseries AEI DBO} & {\bfseries CMP DBO} \\ 
\hline \hline
\multirow[c]{3}{*}{\bfseries Linear Regression} &{$R^2 training$} &0.675	&0.690	&0.666 \\ \cline{2-5} &$R^2 testing$ & 0.777	&0.836	&0.802 \\ \cline{2-5} &$MSE$ &1.591	&1.168	&1.414 \\ \cline{2-5}
\hline \hline
\multirow[c]{3}{*}{\bfseries SVR} &{$R^2 training$} &0.952 &	0.949&	0.952 \\ \cline{2-5} &$R^2 testing$ &0.816	&0.853	&0.817\\ \cline{2-5} &$MSE$ &1.310	&1.045	&1.300 \\ \cline{2-5}
\hline \hline
\multirow[c]{3}{*}{\bfseries Random Forest} &{$R^2 training$} &0.977	&0.978	&0.974 \\ \cline{2-5} &$R^2 testing$ &0.855	&0.876	&0.863 \\ \cline{2-5} &$MSE$ &1.031	&0.881	&0.974 \\ \cline{2-5}
\hline \hline
\multirow[c]{3}{*}{\bfseries Extra Tree Regressor} &{$R^2 training$} &1.000	&1.000	&1.000 \\ \cline{2-5} &\bf{$R^2 testing$} & \bf{0.876}	&\bf{0.891}	&\bf{0.873}\\ \cline{2-5} &\bf{$MSE$} &\bf{0.946}	&\bf{0.779}	&\bf{0.906} \\
\hline
\end{tabular}
\end{table}

\begin{table}[h!]
\centering
\caption{Evaluation results of the four ML models for the vertical AB1 structures dataset.}\label{tab112}
\vspace{5mm}
\begin{tabular}{|c|c|c|c|c|}
\hline 
 &  & {\bfseries ADI DBO} & {\bfseries AEI DBO} & {\bfseries CMP DBO} \\ 
\hline \hline
\multirow[c]{3}{*}{\bfseries Linear Regression} &{$R^2 training$} &0.707	&0.735	&0.687 \\ \cline{2-5} &$R^2 testing$ &0.667	&0.759	&0.668 \\ \cline{2-5} &$MSE$ &2.748	&1.984	&2.736 \\ \cline{2-5}
\hline \hline
\multirow[c]{3}{*}{\bfseries SVR} &{$R^2 training$} &0.973	&0.972	&0.978\\ \cline{2-5} &$R^2 testing$ &0.744	&\bf 0.906	&0.689\\ \cline{2-5} &$MSE$ &
2.108	&\bf 0.772	&2.564 \\ \cline{2-5}
\hline \hline
\multirow[c]{3}{*}{\bfseries Random Forest} &{$R^2 training$} & 0.981	&0.984	&0.978\\ 
\cline{2-5} &$R^2 testing$ &0.834	&0.882	&0.834 \\ \cline{2-5} &$MSE$ &1.367	&0.971	&1.371 \\ \cline{2-5}
\hline \hline
\multirow[c]{3}{*}{\bfseries Extra Tree Regressor} &{$R^2 training$} &1.000	&1.000	&1.000 \\ \cline{2-5} &\bf{$R^2 testing$} & \bf 0.854	&0.894	& \bf 0.853\\ \cline{2-5} &\bf{$MSE$} &\bf 1.203	&0.871	& \bf 1.212 \\
\hline
\end{tabular}
\vspace{5mm}
\end{table}

This is not surprising as the last two models are based on ensemble learning, which aggregates the predictions of a group of decision trees, thus achieving better performance than the individual predictors. Between \textit{Random Forest} and \textit{Extra Tree Regressor}, the latter shows better accuracy with both horizontal and vertical AB1 structures datasets. When comparing the performance of the four models using overlay features from different DBO steps, it is clear that AEI overlay data gives the best accuracy for all models. This is because the overlay data in the AEI step is more stable and has much less outliers than the overlay data from ADI and CMP steps. The outliers are more frequent in ADI because the grating targets in this step still consist of photoresist of M1B layer, and the resist layer is not uniform throughout the whole wafer. Meanwhile in CMP DBO, the measurement is done after the wafer has been metalized and polished, which cause noises in the reflected spectrum.

\begin{figure}
\centering
\captionsetup[subfigure]{labelformat=empty}
    \subfloat[Horizontal structures]{\includegraphics[width=11cm]{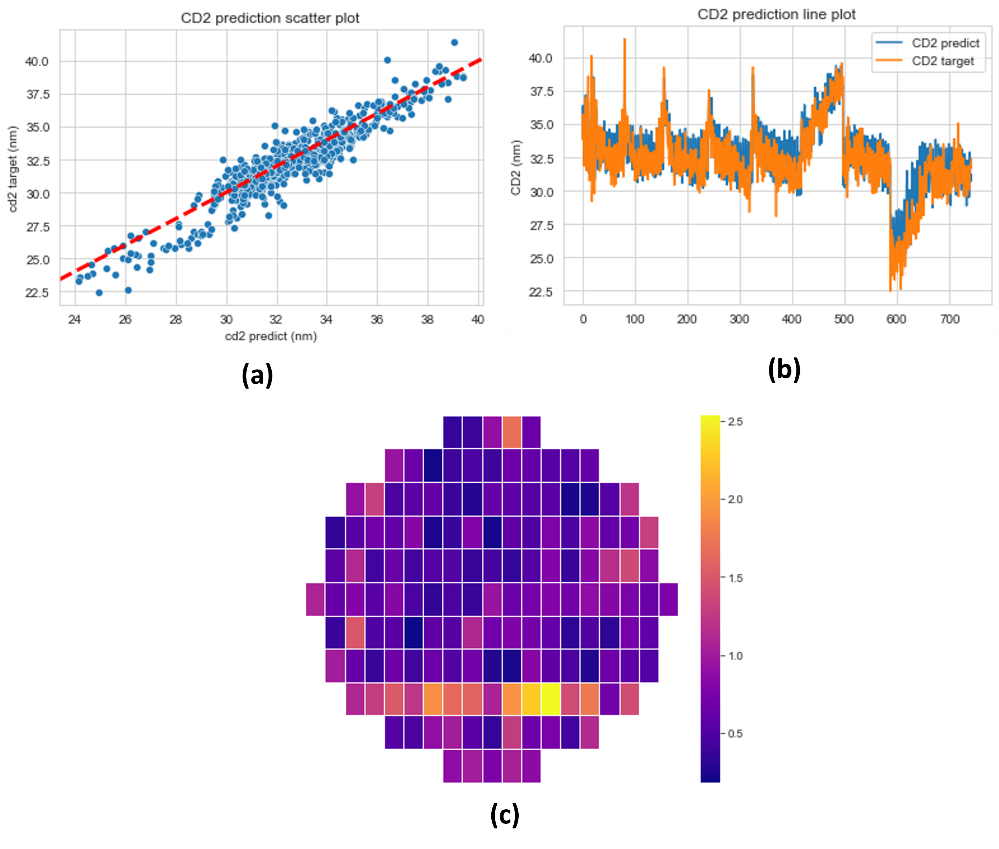} \label{fig53} }
    \qquad
    \subfloat[Vertical structures]{\includegraphics[width=11cm]{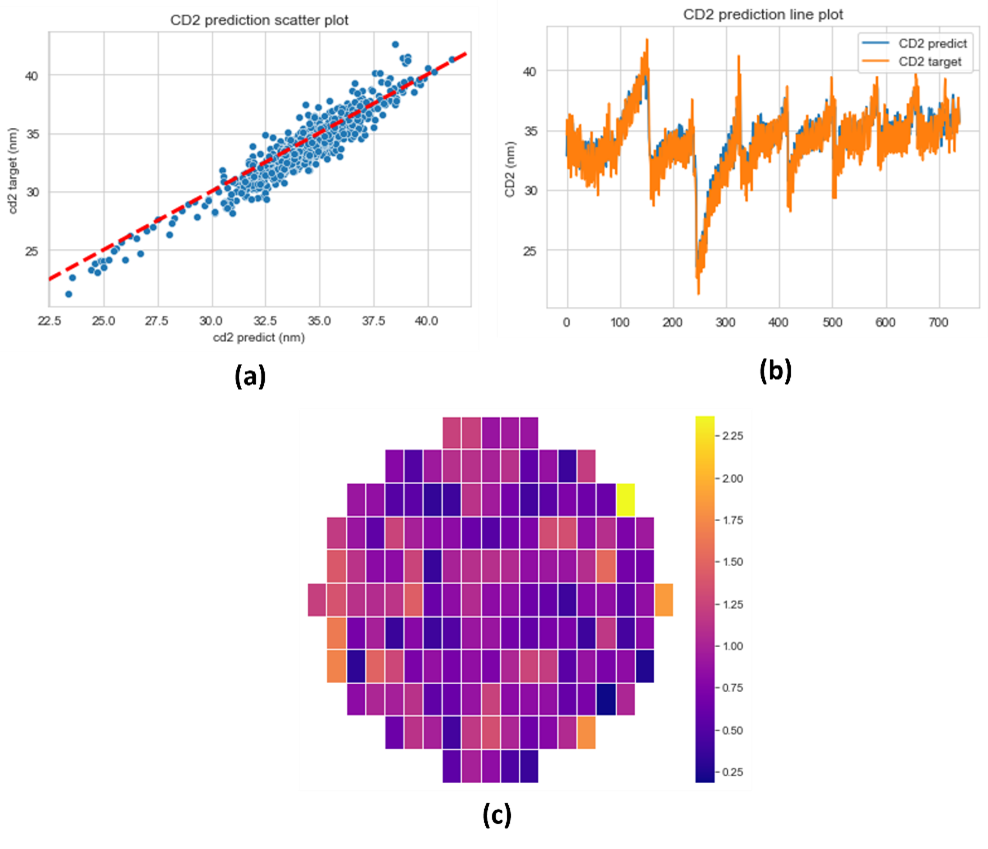} \label{fig54} }
    \caption{Results of \textit{Extra Tree Regressor} using the ADI overlay data to predict CD2\_AB1. (a) CD2\_AB1 prediction scatter plot. (b) Line plots of target and predicted CD2\_AB1 values. (c) Wafer map showing each die’s average prediction error.}
    \label{fig5340}
\end{figure}

To visualize the performance of Machine Learning models in the very early step of the manufacturing process, Fig. \ref{fig5340} shows the results of \textit{Extra Tree Regressor} using the ADI overlay data to predict CD2\_AB1 of horizontal and vertical structures, respectively. Looking at the scatter graphs, the predicted and target values of CD2\_AB1 fits closely to the 45-degree line, which proves that the \textit{Extra Tree Regressor} model performed well on both datasets of horizontal and vertical structures. The well fitted trend is also presented in the line graphs, where the patterns of predicted values are closely mimicked by the patterns of observed data. Finally in the wafer maps, which show the average prediction error of each die, we can see that for the majority of the wafers area, the prediction errors are approximately 1 nm. This error only accounts for 3.125\% of the designed distance between M1A and M1B layers. The highest average prediction error in one die for the horizontal structures is around 2.5  nm (7.813\% of the designed CD2\_AB1), while for the vertical structures it is close to 2.25  nm (7.031\% of the designed CD\_AB1).

\subsection{Capacitance prediction of test structures from overlay data}
After the fabrication process, electrical tests such as capacitance and resistance measurements were conducted on the wafers to check for the quality of fabricated devices. In previous experiment, space-CD measurement data has been successfully predicted from the overlay measurements. This proves that it is also possible to predict capacitance of fabricated structures from overlay errors, as CD-SEM and capacitance are directly correlated. In this section, Machine Learning is used to predict the capacitance of different test structures using overlay measurement data.

\subsubsection{Methodology}
\begin{figure}
\centering
    \subfloat[Wafer D02]{\includegraphics[width=12cm]{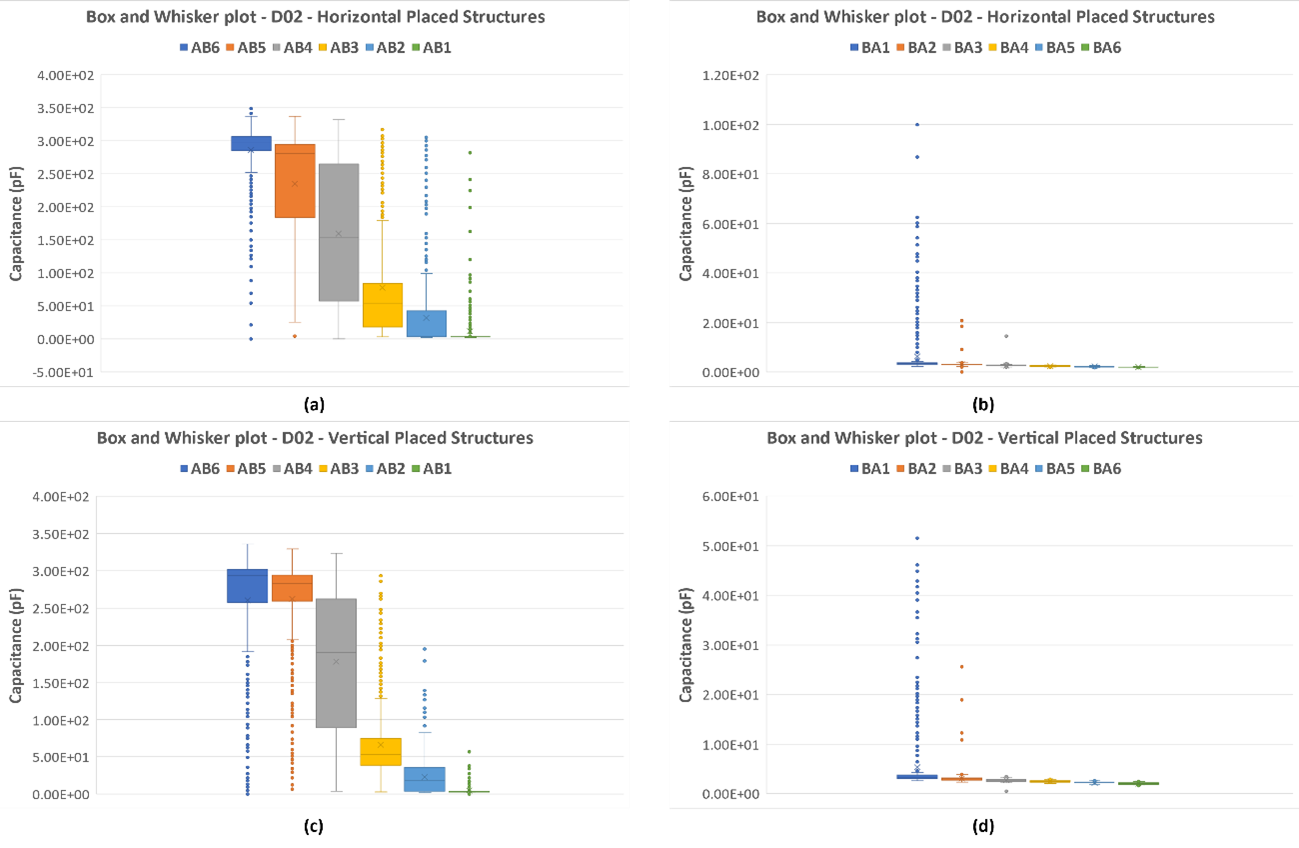} \label{fig55} }
    \qquad
    \subfloat[Wafer D10]{\includegraphics[width=12cm]{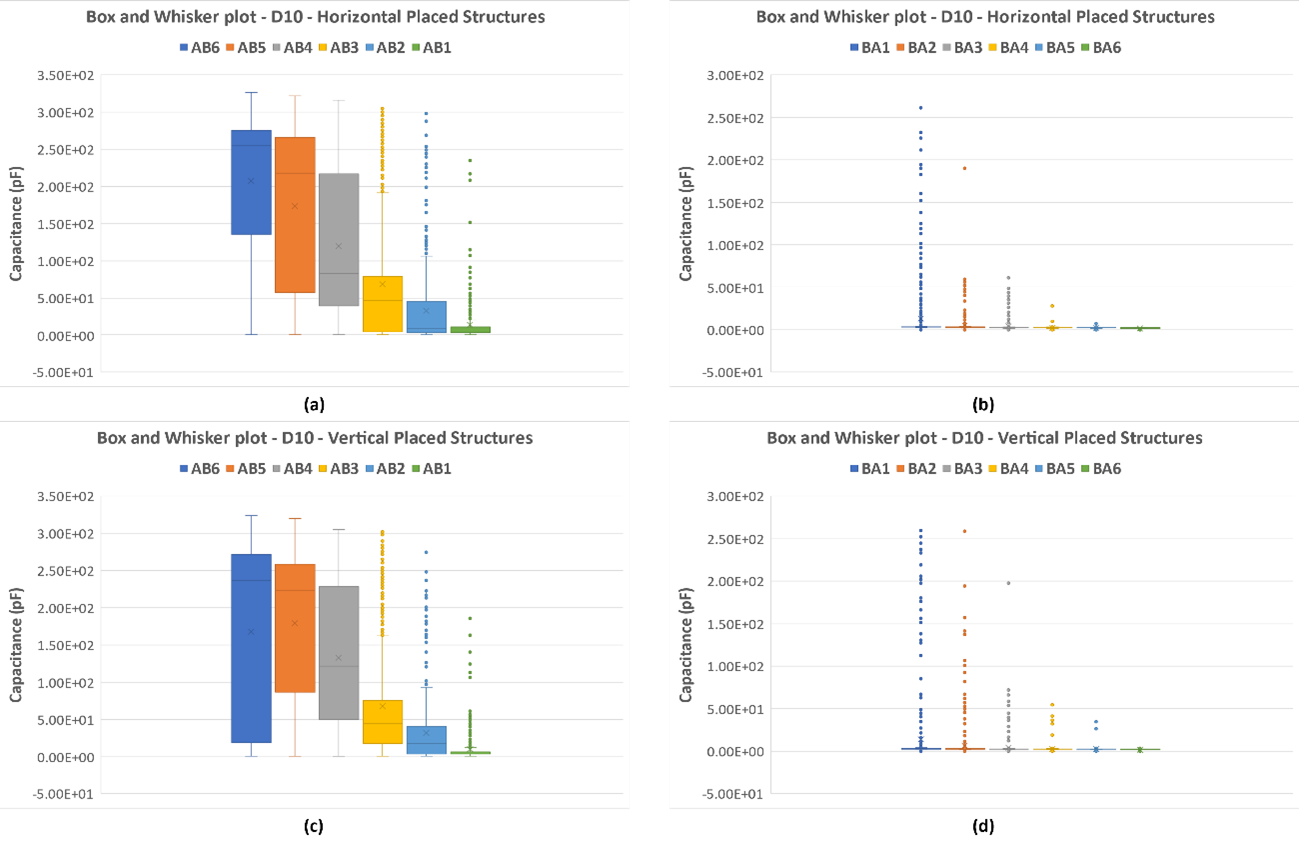} \label{fig56} }
    \caption{Box and Whisker plots of capacitance measurement data for (a) wafer D02 and (b) wafer D10.}
    \label{fig534}
\end{figure}

In this experiment, the capacitance measurement was conducted on two wafers, D02 (Non-Programmed Overlay wafer) and D10 (Programmed Overlay wafer). There are in total 127 dies, and in each five instances of one type of structures (AB1, …, AB6, BA1, …, BA6) were tested. This results in 635 capacitance measurements for each type of test structure in one direction on each wafer. Fig. \ref{fig534} displays the variation in the capacitance measurement data of the two wafers.

In these graphs, the overall capacitance of AB1, AB2, … AB6 structures is greater than that of BA1, BA2, … BA6, which means that the distance between M1A and M1B layers of ABx structures is smaller than BAx structures. The explanation lies in the placement error, i.e., when the M1B layer is shifted to the left by a small distance from the designed position. The small CD2 distance of ABx structures results in many failures and outliers in the capacitance data; moreover, the variance of ABx data is also much larger than the variance of BAx data. Since the target output contains outliers, if we train the regressor model directly with the data, the prediction accuracy will be low as outliers will bias the Machine Learning model during the training. Therefore, before training the model, DBSCAN algorithm was applied to clean the capacitance data. Any abnormal data points detected by the DBSCAN algorithm in one die will be replaced by the mean value of the remained clean data of that die. If there is a die which consists of only outliers, all data points on that die will be replaced by the mean value of the clean data of the whole wafer. The input features fed to the DBSCAN model are measured capacitance and die position in X and Y direction.

DBSCAN (Density-Based Spatial Clustering) is an unsupervised Machine Learning method for special clustering, which views clusters as high-density regions separated by low density areas. Therefore, the clusters discovered by DBSCAN can have any arbitrary shapes, unlike some other clustering algorithms such as k-means, which has an assumption that clusters are in convex shape. The most important hyperparameters of the model are epsilon $\epsilon$, which determines the maximum distance between two samples to be considered neighbours, and \textit{min\_samples}, which specifies the minimum number of samples to form a dense region. The algorithm counts the number of points located within the $\epsilon$ range and if it is at least equal to the \textit{min\_samples} it considers it a core sample. All the datapoints that lie in its $\epsilon$-neighbourhood belong to the same cluster. Datapoints not belonging to any $\epsilon$-neighbourhood of a core sample are considered outliers. In this work, the \textit{min\_samples} parameter is chosen to be 2, the minimum value possible, in order to not replace too many data points because of constraints given by the size of the dataset. To tune the $\epsilon$ parameter, the distance to the nearest point of every data point has been found, sorted and plotted. The optimal $\epsilon$ is the point where the change of the graph is most pronounced. For the regression model to predict capacitance values based on overlay data, there are 3 datasets corresponding to each DBO step, each with 29 input features:

\begin{itemize}
    \item 26 overlay measurements of each die. Overlay error is chosen in X direction if  the capacitance values are from the horizontal structures, and in Y direction if they are from the vertical structures.
    \item Die positions in X and Y directions.
    \item Instances’ indices.
\end{itemize}

Since the data is collected from only two wafers, thus, for each horizontal or vertical test structure, the dataset from two wafers were concatenated, then the training and testing set are divided with a ratio of 80-20. The \textit{Extra Tree Regressor} model was selected for this experiment to predict the capacitance value of different test structures. The reason for this is based on the previous experiment results, \textit{Extra Tree Regressor} performed the best compared to other Machine Learning models that were also tested. The hyperparameters of the model were chosen to be the same as the last experiment, with forest size 60 and minimum sample split size 2. To validate the performance after training, $R^2$ and MSE metrics were used.

\subsubsection{Results and Discussions}
\begin{figure}[h!]
\includegraphics[width=9cm]{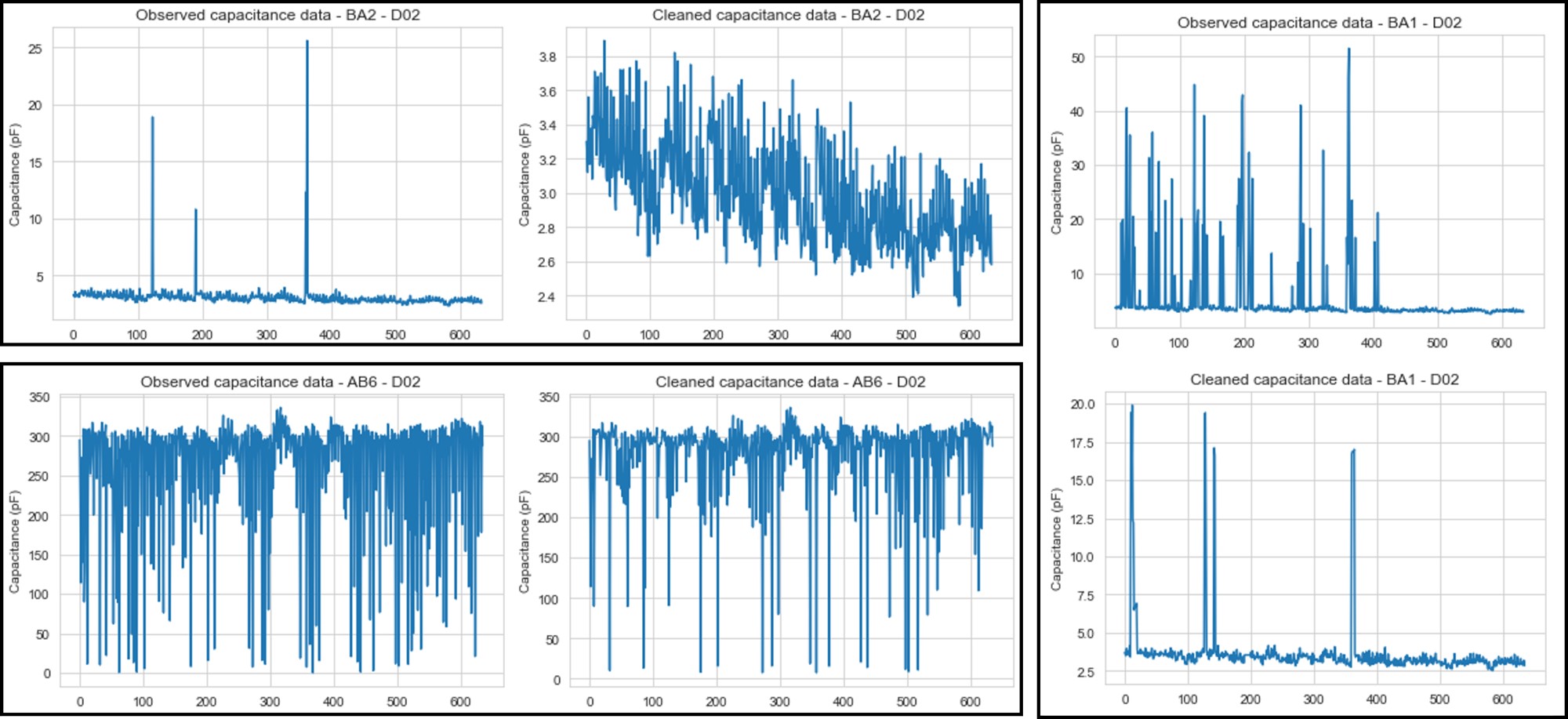}
\centering
\vspace{5mm}
\caption{Examples showing the results of data cleaning with DBSCAN applied to capacitance data of vertically placed BA1, BA2, and AB6 structures from wafer D02.}
\label{fig58}
\end{figure}

Results of applying DBSCAN algorithm to clean measured capacitance data of vertical-placed BA1, BA2, and AB6 test structures from wafer D02 are shown in Fig. \ref{fig58}. From the graphs, we can see that DBSCAN has performed well in the task of cleaning data. For the capacitance measurements of BA2 structures, there are only a few outliers, and the algorithm can spot and replace all of them. This results in the clean dataset with smaller variance compared to the original data. For BA1 structures, there are more failures and abnormalities, yet DBSCAN can detect and clean most of them. However, for AB6 structures, there are a lot of failures, and thus, the data after cleaning is still very noisy. In general, as the gap between M1A and M1B layers of BAx structures is fairly large, capacitance data collected for those structures contains few outliers. Thus, data after applied DBSCAN is clean with most of the outliers removed. Meanwhile, for ABx structures, the distance between two metal layers is small, which causes the failures happening more frequently. Therefore, the capacitance data of ABx structures is very noisy, making it hard for the DBSCAN to clean most of the outliers.

We have conducted an experiment to substantiate the application of DBSCAN algorithm as outlier detection method and if it may negatively affect the performance of the regression models. Table \ref{tab333} shows the prediction results with raw data (without applying DBSCAN) and clean data (after applying DBSCAN). We opted for DBSCAN, as the structures from AB1 to AB6 contain many outliers, which do not really have any related trend with the overlay data. Future work can be extended towards appending other relevant input features along with overlay input to reorient outlier analysis and removal/reduction strategy. After the data cleaning step, the \textit{Extra Tree Regressor} was trained to predict capacitance value of test structures. Table \ref{tab10} and Table \ref{tab11} show the prediction results after training.
\vspace{-0.4cm}

\begin{table}[h!]
\centering
\caption{Comparison of prediction performance of \textit{Extra Tree Regressor} using raw overlay data and clean data after applying DBSCAN for outliers removal.}\label{tab333}
\vspace{2mm}
\begin{tabular}{|p{1.5cm}|p{1.5cm}|p{2cm}||p{1.5cm}|p{1.5cm}|p{2cm}|}
\hline
\multicolumn{3}{|c||}{\bfseries Raw data} & \multicolumn{3}{|c|}{\bfseries Data after data cleaning} \\ \hline \hline 
Device & {\bfseries $R^2$} & {\bfseries $MSE (nF^2)$} & Device & {\bfseries $R^2$} & {\bfseries $MSE (nF^2)$} \\ \cline{1-6}
 \hline \hline
{\bfseries AB6} & 0.612	& 4976.159	& {\bfseries AB6} & 0.524	& 5802.344 \\ \cline{1-6}
\hline
 {\bfseries AB5} & 0.368	& 4697.836	& {\bfseries AB5} & 0.471	& 3668.759 \\ \cline{1-6}
\hline
 {\bfseries AB4} & 0.472	& 4617.266	&  {\bfseries AB4} & 0.588	& 3513.079 \\ \cline{1-6}
\hline
 {\bfseries AB3}&  0.632	& 1300.198	&  {\bfseries AB3}&  0.565	& 1037.417 \\ \cline{1-6}
\hline
 {\bfseries AB2} & 0.582	& 309.921	& {\bfseries AB2} & 0.613	& 260.188 \\ \cline{1-6}
\hline
 {\bfseries AB1} & 0.113	& 145.959	&  {\bfseries AB1} & 0.220	& 42.433 \\ \cline{1-6}
\hline \hline
{\bfseries BA6} & 0.802	&0.006	& {\bfseries BA6} & 0.828	& 0.004 \\ \cline{1-6}
\hline
{\bfseries BA5} & 0.365	& 0.026	& {\bfseries BA5} & 0.823	& 0.006 \\ \cline{1-6}
\hline
{\bfseries BA4} & -0.356	& 0.131	& {\bfseries BA4} & 0.841	& 0.009 \\ \cline{1-6}
\hline
{\bfseries BA3} & 0.679	& 3.579	& {\bfseries BA3} & 0.721	& 2.993 \\ \cline{1-6}
\hline
{\bfseries BA2} & 0.067	& 69.113	&  {\bfseries BA2} & 0.803	& 15.528 \\ \cline{1-6}
\hline
{\bfseries BA1} & 0.742	& 182.100	&  {\bfseries BA1} & 0.950	& 0.077 \\ 
\hline
\end{tabular}
\end{table}

\begin{figure}[h!]
	\begin{center}
		\begin{subfigure}[b]{0.5\textwidth}
		\includegraphics[width=1.1\linewidth]{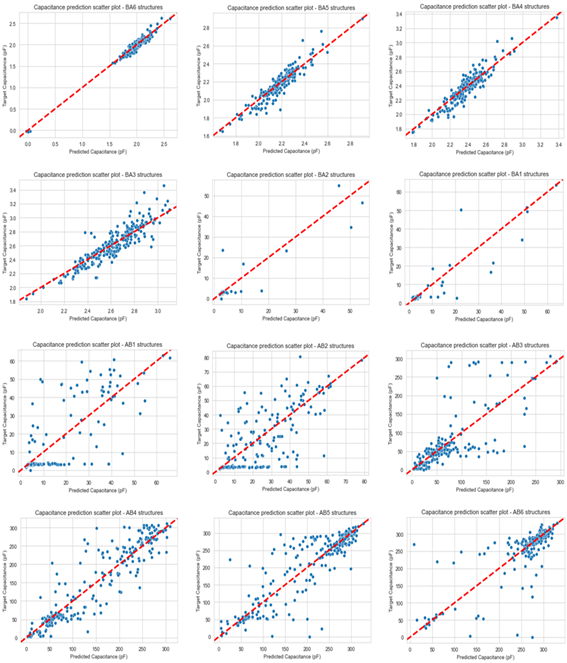}
            \caption{Horizontal structures}
            \label{fig59}
		\end{subfigure}
            \begin{subfigure}[b]{0.5\textwidth}
            \includegraphics[width=1.1\linewidth]{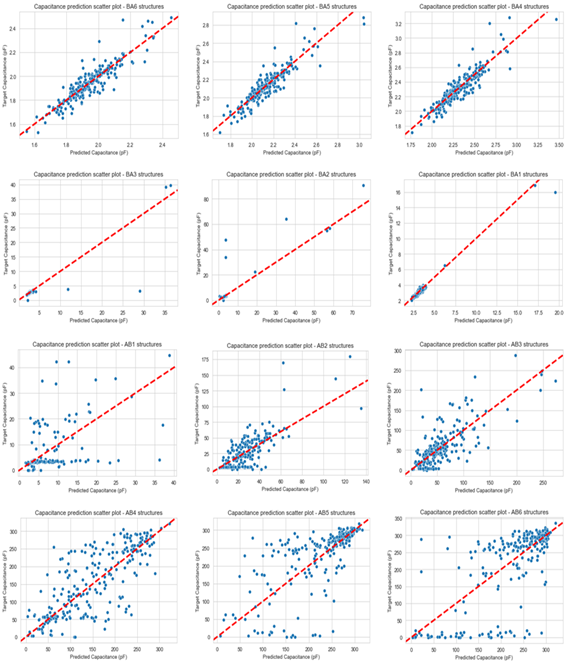}
            \caption{Vertical structures}
            \label{fig510}
            \end{subfigure}
	\end{center}
	\caption 
	{ \label{fig590}
		Scatter plots showing the results of the \textit{Extra Tree Regressor} model when predicting the capacitance value of (a) horizontal-placed structures and (b) vertical-placed structures. }
\end{figure}

It is clear that the trained Machine Learning model performed well with the BAx structures, where the $R^2$ scores of most of the tests are greater than 0.8, while the MSE are small. For the structures BA4, BA5, BA6, the MSE metrics from all the tests are close to 0, whereas for BA1 and BA2 structures, the MSE values are larger, but still smaller than 20 $pF^2$. In contrast, the prediction accuracy of the regressor model is low for the structures of AB1, AB2, …, AB6. In most of the tests, the $R^2$ values are only around 0.6, while the MSE of those tests even goes up to the values of thousands $pF^2$. The reason for the low performance is that the capacitance dataset of ABx structures is very noisy even after cleaning by DBSCAN. Therefore, it is hard for any regression algorithms to achieve decent performance. Comparing the performance of the \textit{Extra Tree Regressor} model when using different DBO data, we can see that the model works best when it was trained and tested by AEI overlay data. While for ADI and CMP overlay data, the performance was slightly decreased. The reason for this has been discussed earlier in the previous section, as the AEI overlay data is more stable than ADI and CMP. The visualizations of the model’s performance on testing dataset are shown in Fig. \ref{fig59} and Fig. \ref{fig510}. The scatter plots in those figures clearly describe the trend of the tables above, since the points spread far away from the $x=y$ line for ABx structures, while they fit closely for BAx structures.

\begin{table}[h!]
\centering
\caption{Examples of performance results of \textit{Extra Tree Regressor} using overlay data from different DBO step when predicting capacitance values of horizontal placed structures.}\label{tab10}
\vspace{2mm}
\begin{tabular}{|c|c|c|c|c|c|c|}
\hline
 \multirow{2}{*}{ Test Structures} & \multicolumn{2}{|c|}{\bfseries ADI DBO} & \multicolumn{2}{|c|}{\bfseries AEI DBO}  & \multicolumn{2}{|c|}{\bfseries CMP DBO} \\ \cline{2-7}
& {\bfseries $R^2$} & {\bfseries $MSE$} & {\bfseries $R^2$} & {\bfseries $MSE$} & {\bfseries $R^2$} & {\bfseries $MSE$} \\ \cline{2-7}
 \hline \hline
 \color{red}
{\bfseries AB6} & \color{red}0.652	& \color{red}1832.295	& \color{red}0.674	& \color{red}1718.611	& \color{red}0.666	& \color{red}1762.366 \\ \cline{2-7}
\hline
 \color{red}{\bfseries AB5} & \color{red}0.784	& \color{red}2075.884	& \color{red}0.778	& \color{red}2136.266	& \color{red}0.778	& \color{red}2141.364 \\ \cline{2-7}
\hline
 \color{red}{\bfseries AB4} & \color{red}0.864	& \color{red}1446.689	& \color{red}0.866	& \color{red}1420.37	& \color{red}0.865	& \color{red}1429.711 \\ \cline{2-7}
\hline
 \color{red}{\bfseries AB3}&  \color{red}0.647	& \color{red}1972.332	& \color{red}0.681	& \color{red}1781.803	& \color{red}0.658	& \color{red}1911.048 \\ \cline{2-7}
\hline
 \color{red}{\bfseries AB2} & \color{red}0.593	& \color{red}157.465	& \color{red}0.615	& \color{red}148.941	& \color{red}0.614	& \color{red}149.283 \\ \cline{2-7}
\hline
 \color{red}{\bfseries AB1} & \color{red}0.658	& \color{red}90.119	& \color{red}0.659	& \color{red}89.752	& \color{red}0.623	& \color{red}99.097 \\ \cline{2-7}
\hline \hline
{\bfseries BA6} &0.964	&0.003	&0.966	&0.003	&0.968	&0.003 \\ \cline{2-7}
\hline
{\bfseries BA5} &0.865	&0.004	&0.82	&0.004	&0.867	&0.004 \\ \cline{2-7}
\hline
{\bfseries BA4} & 0.851	&0.006	&0.826	&0.006	&0.863	&0.006 \\ \cline{2-7}
\hline
{\bfseries BA3} &0.803	&0.014	&0.865	&0.012	&0.815	&0.013 \\ \cline{2-7}
\hline
{\bfseries BA2} & 0.828	&4.458	&0.876	&4.655	&0.805	&5.036 \\ \cline{2-7}
\hline
{\bfseries BA1} & 0.853	&9.586	&0.849	&9.853	&0.806	&12.664 \\ 
\hline
\end{tabular}
\end{table}

\begin{table}[h!]
\centering
\caption{Examples of performance results of \textit{Extra Tree Regressor} using overlay data from different DBO step when predicting capacitance values of vertical placed structures.}\label{tab11}
\vspace{2mm}
\begin{tabular}{|c|c|c|c|c|c|c|}
\hline
 \multirow{2}{*}{ Test Structures} & \multicolumn{2}{|c|}{\bfseries ADI DBO} & \multicolumn{2}{|c|}{\bfseries AEI DBO}  & \multicolumn{2}{|c|}{\bfseries CMP DBO} \\ \cline{2-7}
& {\bfseries $R^2$} & {\bfseries $MSE$} & {\bfseries $R^2$} & {\bfseries $MSE$} & {\bfseries $R^2$} & {\bfseries $MSE$} \\ \cline{2-7}
 \hline \hline
 \color{red}{\bfseries AB6} & \color{red}0.524	& \color{red}5802.344	& \color{red}0.528	& \color{red}5748.09	& \color{red}0.563	& \color{red}5330.72 \\ \cline{2-7}
\hline
 \color{red}{\bfseries AB5} & \color{red}0.471	& \color{red}3668.759	& \color{red}0.417	& \color{red}4044.306	& \color{red}0.457	& \color{red}3768.842 \\ \cline{2-7}
\hline
 \color{red}{\bfseries AB4} & \color{red}0.588	& \color{red}3513.079	& \color{red}0.598	& \color{red}3426.801	& \color{red}0.562	& \color{red}3739.637 \\ \cline{2-7}
\hline
 \color{red}{\bfseries AB3}&  \color{red}0.565	& \color{red}1037.417	& \color{red}0.584	& \color{red}991.975	& \color{red}0.568	& \color{red}1030.288 \\ \cline{2-7}
\hline
 \color{red}{\bfseries AB2} & \color{red}0.613	& \color{red}260.188	& \color{red}0.613	& \color{red}260.644	& \color{red}0.632	& \color{red}247.553 \\ \cline{2-7}
\hline
 \color{red}{\bfseries AB1} & \color{red}0.22	& \color{red}42.433	& \color{red}0.206	& \color{red}43.191	& \color{red}0.179	& \color{red}44.666 \\ \cline{2-7}
\hline \hline
{\bfseries BA6} &0.95	&0.077	&0.951	&0.076	&0.947	&0.082 \\ \cline{2-7}
\hline
{\bfseries BA5} &0.803	&15.528	&0.812	&14.763	&0.812	&14.778 \\ \cline{2-7}
\hline
{\bfseries BA4} &0.721	&2.993	&0.724	&2.967	&0.734	&2.853 \\ \cline{2-7}
\hline
{\bfseries BA3} &0.814	&0.009	&0.828	&0.009	&0.792	&0.01 \\ \cline{2-7}
\hline
{\bfseries BA2} & 0.823	&0.006	&0.836	&0.006	&0.798	&0.007 \\ \cline{2-7}
\hline
{\bfseries BA1} & 0.828	&0.004	&0.832	&0.004	&0.791	&0.005 \\ 
\hline 
\end{tabular}
\end{table}

\clearpage
\newpage
\section{Conclusions}
This work aims to analyse the overlay data in the semiconductor manufacturing and to make use of the overlay measurements from early steps in the process to predict electrical property of the final fabricated structures using machine learning techniques. Several data-driven techniques were developed to predict critical dimension measurements from overlay metrology data, in particular, the distance between two metal layers of the test structures. The prediction accuracy of all models is reasonable, with \textit{Extra Tree Regressor} being the best performing algorithm, as the $R^2$ values in all cases were close to 0.9 and the $MSE$ values were close to 0. We demonstrated the applicability of appropriate machine learning models to improve the throughput, as now, the critical dimensions of every wafer can be predicted from the overlay metrology data, which reduces the number of wafers needed to be measured by the physical CD-SEM tools. Future motivation is to include defective test structures (with short circuits or line breaks) as well as larger datasets (measuring more wafers from different lots from well optimized process steps) to improve the predictive performance. Furthermore, to consider including other metrology data, such as mask measurement data, resist thickness, line width roughness, etc. and further optimization using hyperparameters tuning techniques such as Grid Search, Random Search, or Bayesian Optimization are our future directives.

\clearpage
\printbibliography

\end{document}